\documentclass[prl,aps,tightenlines,twocolumn]{revtex4}
\usepackage[]{times,amsmath,epsfig,graphicx}
\begin{document}

\title{Passively mode locked Raman laser}

\author{W. Liang, V. S. Ilchenko, A. A. Savchenkov, A. B. Matsko, D. Seidel, and L. Maleki}

\affiliation{OEwaves Inc., 2555 E. Colorado Blvd., Ste. 400, Pasadena, California 91107}

\begin{abstract}
We report on the observation of a novel mode locked optical comb generated at the Raman offset (Raman comb) in an optically pumped  crystalline whispering gallery mode resonator. Mode locking is confirmed via measurement of the radio-frequency beat note produced by the optical comb on a fast photodiode. Neither the conventional Kerr comb nor hyper-parametric oscillation is observed when the Raman comb is present.
\end{abstract}


\maketitle

Monolithic whispering gallery mode (WGM) resonators provide numerous opportunities to study fundamental physical phenomena in transparent optical materials. Ultra-low threshold stimulated Raman scattering (SRS) and four-wave mixing (FWM) \cite{spillane02n} resulting in the generation of a novel optical frequency (Kerr) comb  \cite{delhaye07n} are among these phenomena. Kerr combs have rapidly become the subject of extensive research \cite{delhaye07n,delhaye08prl,savchenkov08prl,grudinin09ol,agha09oe,levy09np,razzari09np,braje09prl,delhaye09arch,arcizet09chapt,matsko09sfsm,chembo10prl} because the excellent uniformity of the comb lines as well as their high frequency of their repetition rate makes them interesting for many practical applications currently served by the more conventional combs produced with femtosecond mode-locked lasers. While generation of the Kerr comb and multi-frequency SRS are independent phenomena, in this paper we describe results of an experimental study to show that  generation of a mode locked comb at Raman frequency offset (Raman comb) is feasible. We argue that this is a novel regime of mode locking.

Generation of a Kerr comb is based on nonlinear processes taking place in optical media possessing cubic nonlinearities \cite{kippenberg04prl,savchenkov04prl}. The comb starts from  hyper-parametric transformation of two pump photons into  signal and idler photons with frequencies different from the pump frequency  ($2\hbar \omega_p \rightarrow \omega_s+\omega_i$), and then broadens because of the FWM among the photons.  WGM resonators are ideally suited for observation of the process as they lead to extremely high continuous wave (cw) optical intensity within the nonlinear media, which results in a reduction of the threshold of the nonlinear process with respect to the external cw pump power.

Hyper-parametric oscillation in WGM resonators has very rich dynamics. The oscillation is observed as generation of optical sidebands equally separated from the pump frequency. Usually, only one pair of signal and idler sidebands is generated with lower powers of the external pump \cite{savchenkov04prl}. An increase of the power leads to the increase of  number of sidebands and eventual growth of an octave-spanning optical frequency comb \cite{delhaye09arch}. The comb usually grows from the inside, but the conventional scenario for comb generation frequently does not work. For example, depending on  resonator parameters, the first pair of  sidebands can be generated about a THz from the pump frequency, with subsequent population of comb lines growing from sidebands to the carrier \cite{savchenkov08prl,chembo10prl}. The comb is usually symmetric, but a well developed comb becomes asymmetric with respect to the pump frequency \cite{delhaye07n,savchenkov08prl}, and the comb repetition rate becomes power dependent and different from the free spectral range of the WGM resonator.

In spite of significant ongoing research, to our knowledge there is no complete understanding of the Kerr comb evolution. It was pointed out \cite{matsko09sfsm} that the Kerr comb resembles an additive modulational instability in ring laser studied two decades ago \cite{haelterman92ol,coen97prl,coen01ol}. According to these earlier studies,  modulational instability-based lasing occurs for both positive and negative dispersion when light is confined in a resonator. Results of those studies were adopted for description of a WGM hyperparametric oscillator \cite{matsko05pra} with the conclusion that  oscillation is possible and stable for any group velocity dispersion created within the resonator. Later it was suggested that  negative group velocity dispersion is preferred for  oscillation \cite{agha09oe}. Finally, the conclusion that generation of a comb is unstable at zero group velocity dispersion was made in \cite{chembo10prl}. An attempt to predict the shape of the generated optical pulse by solving a forced nonlinear Schrodinger equation was recently presented \cite{matsko09sfsm}. However, observed pulses are significantly longer than than the theory predicts \cite{arcizet09chapt}. The  experimental study reported here shows that the Kerr comb is preferentially generated in the region of {\em normal} group velocity dispersion; this calls for further research in the field.

SRS is one of the processes hindering the generation of a Kerr comb. SRS was studied since early days of WGM resonators in liquid droplets \cite{qian85ol,lin94prl}. SRS in very high-Q WGM resonators attracted attention of researchers somewhat earlier than hyper-parametric oscillation did \cite{spillane02n}. It was shown that the Raman process starts with low pump powers, at the microwatt level, and that the efficiency of  frequency conversion is very high \cite{kippenberg04jstqe,grudinin07ol}. The possibility of generation of numerous higher order Stokes lines was demonstrated \cite{grudinin07ol}.

Four wave mixing has nearly the same efficiency as SRS, so that hyper-parametric oscillation and Raman lasing have nearly identical thresholds in WGM resonators \cite{lin94prl,matsko05pra}. SRS tends to have lower threshold in larger resonators.  This is because, in contrast to FWM, lasing process does not require phase matching, and so Stokes light is generated in an arbitrary mode having a higher quality factor and better overlap with the pumping mode. Stokes lines may involve several WGMs. Light generated in each mode represents an independent laser, so there is {\em no coherence} among the modes. If  SRS starts at a lower pump power, the harmonics of the Stokes component fold down to the frequency of the pump, forming optical sidebands resembling those produced by hyper-parametric oscillation. Since the Stokes signal and the pump are mutually independent,  downconverted sidebands and the pump have low mutual coherence. The SRS process diminishes the quality of the optical comb produced with low enough pump power \cite{savchenkov08oe}. Higher pump power can result in the overlap of Kerr and Raman combs, and in generation of a joint mode locked comb \cite{delhaye07n}. In some regimes, though, SRS results in an appreciable nonlinear loss for the pump radiation and drains the pump so fast that no Kerr comb is formed at any pump power \cite{grudinin07ol}. Therefore, to study the dynamics of Kerr combs one needs to suppress the SRS process.

We have discovered that  SRS lines generated in a calcium fluoride WGM resonators can be self mode locked. The Stokes light generated in such resonators has multiple equidistant optical harmonics that cannot be considered as mutually incoherent. The phase locking of  Stokes modes occurs for specific values of dispersion, nonlinearity, and the width of the Raman gain. In such a regime, no high order Stokes modes are generated, and no hyper-parametric oscillation is observed. To confirm that  mode locking occurs we have measured  spectral purity and  instantaneous linewidth of the RF signal generated by the beat of the Raman comb on a fast photodiode. We have directly observed the collapsing the linewidth of the RF signal when the mode locking regime was achieved. The low phase noise of the RF signal is indicative of  the high efficiency of mode locking in the Raman comb.

We explain these observations using the notion of passive mode locking developed several decades ago \cite{martinez84ol,haus86jqe,haus00jstqe}. Mode locking occurs by  proper interplay between dispersion and nonlinearity in materials that possess optical (Raman, in our case) gain. The mode locking phenomenon reported here is different from the effect studied in lumped systems, since gain, nonlinearity and dispersion are truly distributed in our case. The physics behind the observed phenomenon is also different from the physics of Kerr combs, as the phase of the generated light is independent of the phase of the pumping light in this mode-locked Raman laser. Moreover, the detuning of the pump frequency from the corresponding WGM eigenfrequency ensures that  phase matching condition necessary for hyper-parametric oscillation occurs \cite{matsko05pra}; while the Raman modes are, to the first approximation, generated at frequencies independent from the pump laser frequency.

Our observations pave the road for the development of a new class of mode locked lasers based on cw pumped WGM resonators. The fundamental advantage of these lasers draws from the extremely high Q-factors of modes of the active optical resonator that ensure high quality mode locking; and the high stability of lasing  is not dependent on the quality of the laser used for pumping. A practical advantage of mode locked Raman lasers is their small size and low power consumption.

\begin{figure}[ht]
\centerline{\epsfig{file=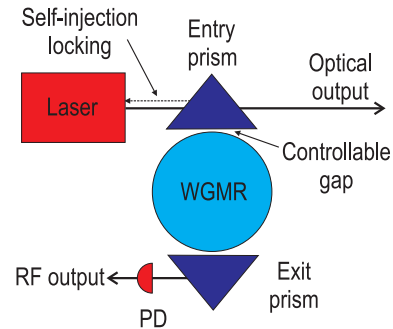,width=5.0cm,angle=0}}
\caption{\label{figure1} Schematic of the experimental setup. Light from the pump laser enters the whispering gallery mode resonator (WGMR) through the entry prism. The light exiting the entry prism is collimated and sent to an optical spectrum analyzer. The output light from the exit prism is sent to a fast photodiode (PD).}
\end{figure}

To realize a WGM mode-locked Raman laser one needs a pumping laser, a WGM resonator made out of a transparent material possessing cubic nonlinearity, an evanescent field coupling element, and supportive optics. We have used a DFB laser, a CaF$_2$ resonator, and glass prism couplers (Fig.~\ref{figure1}). The pump laser produced 6~mW of power at 1532~nm, and we were able to introduce about 2~mW into the WGM resonator. The resonator had 1.9~mm diameter and 0.1~mm thickness. The rim of the resonator was shaped  to maximize optical coupling \cite{maleki09chapter}. The unloaded Q-factor of the resonator was larger than $ 10^9$, and loading with the prisms reduced the Q-factor value to $5 \times 10^8$. We were able to monitor and change the degree of loading of the WGM by changing the distance between the coupling prisms and the resonator surface. The laser was self-injection locked to a selected WGM. Injection locking reduced the 2~MHz linewidth of the free running DFB laser to $\approx 300$~Hz \cite{maleki09chapter}, which allowed us to pump the high-Q WGM efficiently.

\begin{figure}[ht]
\centerline{\epsfig{file=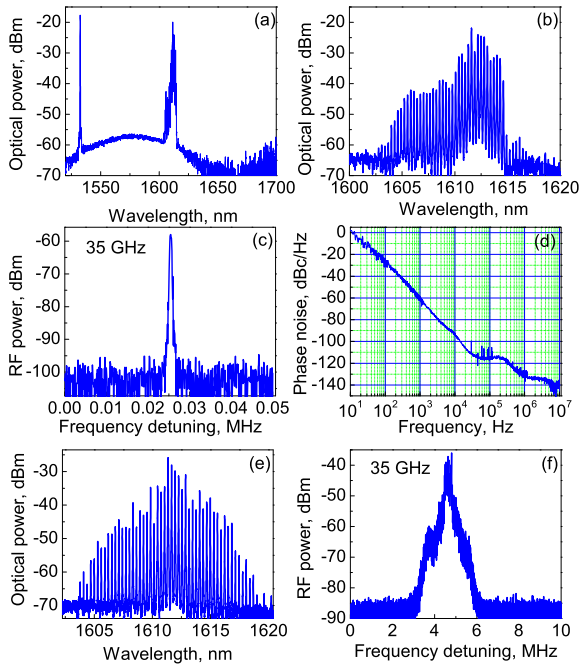,width=8.5cm,angle=0}}
\caption{\label{figure2}  Optical frequency comb formed at the first Stokes line of the CaF$_2$ WGM resonator.  Neither the second Stokes line (expected at 1699~nm) nor the FWM sidebands around the carrier are observed. (a) Optical spectrum at the entry prism. (b) The comb has approximately 10~nm spectral width. (c) Spectrum of the RF signal generated by the Raman comb on a fast photodiode taken with an RF spectrum analyzer. The resolution bandwidth is 50~kHz. (d) Single sideband phase noise of the RF signal. (e) An example of the optical spectrum of the incoherent Raman comb. (f) RF signal generated by demodulating the incoherent comb on a fast photodiode.}
\end{figure}

With an optimal value of the optical power and optimal tuning of the pumping laser we achieved generation of the Raman comb shown in  Fig.~(\ref{figure2}a and \ref{figure2}b). The shape of this Raman frequency comb is similar to the shape of the Raman comb  observed in silica microspheres \cite{spillane02n}. Interestingly, there are no satellite sidebands generated around the carrier frequency of the pumping mode, as was the case with the silica resonator. By varying the pumping frequency with respect to the corresponding WGM it is possible to generate minor FWM-mediated sidebands; however, these sidebands have at least 30~dB lower power as compared with power of the pump. In addition, we did not observe the second order Stokes line, expected at 1699~nm. These facts indicate that the first order SRS process is very efficient. To confirm this, we estimated, from Fig.~(\ref{figure2}), that the power of the Stokes light emitted in the forward direction through the entry prism is approximately $-16$~dBm (we do not collect all the light escaping the entry prism so the number is good for comparison only), so the total power emitted in the Raman modes through the entry prism (both directions) is $-13$~dBm. To compare, the pump power exiting the resonator through the entry prism is $-18$~dBm. Hence, the efficiency of the Raman laser exceeds 50\%.

We studied the RF signal generated by the Raman comb with a fast photodiode and observed the collapse of the linewidth at certain detuning of the pump laser lock point frequency. The spectrum of generated RF signal is shown in Fig.~(\ref{figure2}c). The instantaneous linewidth of the signal is less than 100~Hz. We measured the phase noise of the signal and found that it is quite small (Fig.~\ref{figure2}d). These observations confirm that  modes in the Raman comb are mutually coherent, and that  this Raman laser operates in mode locked condition. This is the basic result of the reported research.

The mode locking regime is not generic in the system. The best results are obtained with pumping critically coupled modes belonging to the basic WGM family. No mode locking is observed if the pump power, and the Q-factor of  WGMs, is not optimal Fig.~(\ref{figure2}e). Even though the optical spectrum of the unlocked Raman comb, Fig.~(\ref{figure2}e), looks nearly the same as the spectrum of the mode locked comb, Fig.~(\ref{figure2}b), the RF signal generated at the fast photodiode, Fig.~(\ref{figure2}f) is much wider compared with the signal generated by the mode locked comb, Fig.~(\ref{figure2}c). This implies that modes of the Raman comb shown in Fig.~(\ref{figure2}e) are phase independent.

\begin{figure}[ht]
\centerline{\epsfig{file=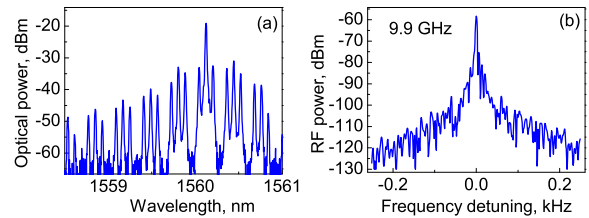,width=8.5cm,angle=0}}
\caption{\label{figure3}  An example of mode locked Kerr comb. (a) Optical spectrum of the comb. (b) RF signal generated by demodulating the comb on a fast photodiode. The resolution bandwidth is 3~Hz.}
\end{figure}

To highlight the difference between the mode locked Raman lasing and the Kerr comb generation  we repeated the measurements with a 1560~nm DFB laser and a WGM resonator with 9.9~GHz FSR using identical setup. We did not observe Stokes modes in that experiment. Instead, the classical Kerr comb was generated around the pumping frequency (Fig.~\ref{figure3}a). The light exiting the resonator produced an excellent RF signal at 9.9~GHz (Fig.~\ref{figure3}b), which confirms mode locking of the comb. The Kerr comb was more stable compared with the Raman comb, so we were able to reduce the resolution bandwidth of the measurement to 3~Hz (Fig.~\ref{figure3}b).

To summarize, our experiments have shown that pumping a 1.9~mm CaF$_2$  WGM resonator with 1532~nm light results in generation of Raman comb around 1610~nm. The Raman comb can be mode locked, and no Kerr comb is generally observed. On the other hand, pumping a 6.8~mm CaF$_2$ resonator with 1560~nm light results in preferential generation of a mode locked Kerr comb with no SRS observed.

\begin{figure}[ht]
\centerline{\epsfig{file=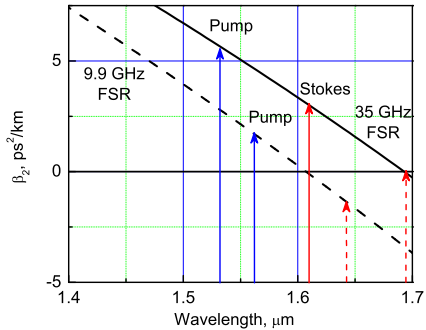,width=6.5cm,angle=0}}
\caption{\label{figure4}  Plot of the calculated group velocity dispersion for  resonators  used in the experiment. Blue solid arrows show the wavelength of the pump used. The solid red arrow shows the position of the Raman comb observed. Dashed red arrows show the position of expected, but not observed, Raman combs.}
\end{figure}

To understand these observations we recall that the group velocity dispersion for different sized resonators is different because of the contribution of  geometrical dispersion of WGMs \cite{matsko03josab}. Plotting the resultant dispersion of  resonators using  Sellmeier's dispersion equation \cite{daimon02ao} for  bulk material, together with the expression for geometrical dispersion of the basic sequence of WGMs \cite{matsko03josab} we find that combs tend to appear at small {\em normal } group velocity dispersion (Fig.~\ref{figure4}). The first Stokes sideband is expected in the region of anomalous dispersion in the 6.8~mm resonator. However, no generation of Raman comb occurs, and a Kerr comb is generated instead. A similar observation is related to the smaller resonator, i.e. the second order Stokes signal is expected in the region of zero dispersion of the resonator, however the sideband is not generated. These observations call for an extensive theoretical study of Kerr combs, since the theory predicts that the region with anomalous group velocity dispersion is preferable for comb generation \cite{agha09oe}. The optical properties of  the resonator material and the geometrical dispersion of arbitrarily shaped WGMs also should be studied.

Generation of the Kerr comb and SRS in a WGM resonator depends on quality factors of  modes, on detuning of the locking point of the laser, and on the laser power. Our experimental setup does not allow us to change these parameters independently: modifying the value of the laser current changes the power and the frequency of the laser, and changing the loading of WGMs changes  properties of the injection lock, and so on. This prohibits us from finding the exact relationship between parameters of the setup and  conditions corresponding to a particular process observed in the resonator. Nonetheless, we have definitely found that smaller resonators tend to generate Raman combs, both locked and unlocked. Moreover, by tuning the loading of the resonator and selecting deeper WGMs possessing different geometrical dispersion, we were able to generate a stand alone Kerr comb. On the other hand, we were unable to generate a Raman comb in  a larger resonator pumped with 1560~nm light.

Our observation of the mode locked Raman combs can again be explained within the framework of  theory developed for passive mode locked lasers \cite{martinez84ol,haus86jqe,haus00jstqe}, where it was shown that mode locking occurs for laser loops with both normal and anomalous dispersion. The optimal operation of
the laser takes place in the vicinity of zero group velocity dispersion, similar to our case. However, the transformation of the theory of mode locked lasers to the Raman comb is not straightforward since Raman gain does not possess saturation properties used in those models. In our case, saturation occurs with draining of the pump power, and the Raman gain stays unchanged with respect to Stokes modes \cite{matsko06oc}.  The description for operation of mode locked Raman lasers will be analyzed theoretically elsewhere.

To conclude, we have experimentally demonstrated a mode locked optical Raman comb in a CaF$_2$ WGM resonator. The comb manifests itself when the frequency of the Stokes Raman line approaches, but does not overlap with, the zero-group velocity frequency region of the WGM spectrum. Upon demodulation on a fast photodiode, the comb generates a spectrally pure RF signal at a frequency corresponding to the repetition rate of the optical comb. This comb is different from the Kerr comb and from passively mode locked lasers, and clearly represents another physical phenomenon uniquely observable in high-Q WGM resonators.

The authors acknowledge Danny Eliyahu's help with the phase noise measurement of the RF signal.

%%%%%%%%%%%%%%%%%%%%%%%%%%%%%%%%%%%%%%%%%%%%%%%%%%%%%%%%%%%%%%%%
\bibliographystyle{IEEEtran}

%%%%%%%%%%%%%%%%%%%%%%%%%%%%%%%%%%%%%%%%%%

\end{document}